\documentclass[aip,jcp,preprint,twocolumn]{revtex4-1}

\usepackage{amsmath}
\usepackage{amsfonts}
\usepackage{amssymb}
\usepackage{graphicx}

\begin{document}

\title{
   Predicting highly correlated hydride-ion diffusion in SrTiO$_3$ crystals
   based on the fragment kinetic Monte Carlo method with machine-learning potential}

\author{Hiroya Nakata}
\email{hiroya.nakata.gt@kyocera.jp}
\affiliation{Kyocera Corporation, Research Institute for Advanced Materials and Devices, 3-5-3 Hikaridai Seika-cho Soraku-gun Kyoto 619-0237, Japan.}

\begin{abstract}
   Oxyhydrides have drawn attention because of their fast ion conductivity 
   and strong reducing properties. 
   Recently, hydride ion migration in SrTiO$_{3-x}$H$_{x}$  oxyhydride crystals 
   has been investigated,  showing that
   hydride ion migration is blocked by slow oxygen diffusion.
   In this study, we investigate the hydride-ion migration mechanism using a kinetic Monte Carlo approach
   to understanding the relationship between the hydride and oxygen ions.  
   The difficulties in applying the  method to hydride and oxygen ion migration involve complex changes in the ionic migration barrier, 
   which shifts dynamically depending on the characteristics of the surrounding hydride and oxygen ions.
   We can predict these complex changes using 
   a machine-learning neural network model. The simulation can then be performed 
   using this model to predict the temperature-dependent ionic-migration behavior.
   We found that our simulation results with respect to the activation barrier for hydride ion diffusion accorded well with those obtained by experiment.
   We also found that hydride ion migration
   is affected by slow oxygen diffusion and that oxygen diffusion
   is accelerated by changes in the ionic migration barriers.  
   The parallel-processing efficiency of our proposed method was 84.92 \% for our 1,000-CPU implementation, suggesting
   that the approach should be widely applicable to simulations of ionic migration in crystals
   at a reasonable computational cost.
\end{abstract}


\maketitle



\section{Introduction}

   Ionic conductors are attracting great interest 
   because of their potential use in many electronic devices such 
   as batteries\cite{zhao2012hierarchical,suntivich2011design,kamaya2011lithium,malavasi2010oxide},
   superconductors\cite{kamihara2008iron},
   photocatalytic devices\cite{litter1999heterogeneous},
   and solid fuel cells\cite{ishihara2009perovskite,malavasi2010oxide}. 
   Recently, a number of studies have investigated
   replacing a significant proportion of the hydride ions by oxide ions to
   form
oxyhydrides\cite{janotti2007hydrogen,koch2012photoconductive,hayashi2002light,bouilly2015electrical,
   yajima2012epitaxial,kobayashi2012oxyhydride,tassel2014direct,tassel2016high}. 
   Because  
   of their fast ion conductivity and strong reducing properties, oxyhydrides have potential as effective charge carriers in traditional electrochemical  
   applications\cite{zhang2017anion,yamaguchi2016large,kobayashi2016pure,yamazaki2013proton},
   and as catalysts for ammonia synthesis\cite{kobayashi2017titanium}.
    
   Significant efforts have been made toward understanding the experimental results for oxyhydride materials.
   The mobility of hydride ions has been measured for LaSrCoO$_3$H$_{0.70}$ via
   quasi-elastic neutron scattering\cite{bridges2006observation} and
   for La$_{2xy}$Sr$_{x+y}$LiH$_{1x+y}$O$_{3y}$ via impedance measurements.\cite{kobayashi2016pure}
   The kinetics of  hydride ions in ABO$_3$-type perovskite materials  are particularly  challenging 
   because the materials have high conductivity, making impedance measurements difficult to apply. 

   Hydride ion mobility in SrTiFeO$_3$ has been measured\cite{steinsvik2001hydrogen}
   but the experiments could not separate fully the hydride ion diffusion from the oxide anion diffusion,
   and the kinetics of the hydride ions were unclear. Hydride ion diffusion in ATiO$_{3-x}$H$_x$
   could be observed via the exchange reactions with D\cite{kobayashi2012oxyhydride},  N\cite{yajima2015labile,masuda2015hydride},
   and F\cite{masuda2015hydride}. 
   The kinetics of H/D exchange have also been analyzed by quadrupole mass spectrometry\cite{tang2017hydride}.
   
   Recently, Liu et al.\cite{liu2019highly} measured the self-diffusion coefficient for hydride ions 
   by combining secondary ion mass spectrometry and 
   a first-principles study, suggesting an activation barrier of around 0.30 eV. 
   Although this experimental study partially explained hydride ion diffusion,
   both experimental and theoretical studies on this topic have been limited
   in comparison to those dealing with the diffusion of oxide anions.  
   Oxygen migration in oxyhydride materials has also not been investigated.

   Theoretical approaches also have powerful tools for understanding hydride ion diffusion. 
   Iwazaki et al.\cite{iwazaki2010negatively,iwazaki2014diversity} investigated the complex-defect electronic 
   structure of H in  BaTiO$_{3-\delta}$ and SrTiO$_{3-\delta}$, and
   with the simulation results suggesting that  the hydride ions were trapped by V$_O$.
   Liu et al.\cite{liu2017formation,liu2018formation} investigated the formation and migration energy of H in 
   the BaTiO$_{3-\delta}$ and K$_2$NiF$_4$ types of oxyhydrides, which revealed the electronic configuration of the hydride ions. 
   This first-principles study suggested that the activation barrier for hydride ion diffusion is 0.17 eV for SrTiO$_{3-\delta}$\cite{liu2019highly} 
   and  0.28 eV for BaTiO$_{3-\delta}$\cite{liu2018formation}. 
   Because the simulated activation barrier for hydride ions (0.17 eV) was less than that obtained by experiment (0.30 eV),
   it is suggested that the fast hydride ion migration is retarded by slow oxide anion diffusion. 
   Although theoretical first-principles studies suggest that slow oxide anion diffusion can retard
   hydride ion migration, how the presence of oxide anions affects the kinetics of hydride ion diffusion 
   is still unclear.  Therefore, kinetic simulation of oxyhydride diffusion remains an interesting subject of investigation.

   The kinetic Monte Carlo (kMC) method\cite{kMCbook,kMCbasis01,kMCbasis02} can be used to predict long-time-scale ionic 
   diffusion in crystals, where the ions jump into the nearest site according to experimentally obtained or simulated 
   activation barriers. The kMC method has been used to predict the kinetics of many types of ionic migration and chemical reactions
   such as yttria-stabilized zirconia\cite{lau2009kinetic,lau2008kinetic},
   oxygen diffusion in doped ceria\cite{ceria01,ceria02},
   various chemical reactions\cite{kMCCO01,kMCN01,kMCwat01,KMCwat02,kMCNH301,kMCNH302,kMCMeOH},
   material diffusion\cite{kmcOld01,matera2011adlayer,temel2007does,rieger2008effect,kMCSOFC},
   electrochemical impedance\cite{pornp2007electrochemical},
   chemical catalysis\cite{hansen1999modeling,hansen2000first,hansen2000first2,reuter2006first,stamatakis2011first,boscoboinik2008monte,kunz2015kinetic},
   and crystal growth\cite{gilmer1980computer,schulze2004hybrid}.

   Recently, we proposed an efficient parallel-processing approach for predicting ionic diffusion with fragment-based kMC (FkMC)\cite{NAKATA2020109844}.
   The method has been applied to SrTiO$_{3-\delta}$ systems using a simple Manning model\cite{manning1968diffusion} as a pilot test.
   The activation barrier of hydride ions found\cite{NAKATA2020109844} did not accord with experimentally derived 
   activation barriers\cite{liu2019highly}, suggesting that the simple Manning model is inadequate for predicting ionic diffusion behavior in oxyhydrides.
   A major reason for the discrepancy may be attributed to inhomogeneous potentials, i.e., the previous simulation model
   did not consider the complex potential energy surface (PES) composed of both hydride and oxide ions.   
   Inclusion of such a complex oxyhydride PES into the kMC method is not straightforward,
   making the kinetic analysis of oxyhydrides a challenging issue in terms of computational science. 

   The aim of this study is to develop a kMC model that can simulate complex ionic migration in crystals, 
   with the method being applied to predicting the kinetics of hydride and oxide ions in SrTiO$_{3-\delta}$ oxyhydrides. 
   To achieve this, the activation barrier for each hydride and oxide ion is refined based on changes in potential energy. 

   Such a refinement of the activation barrier using potential energy differences has been proposed by Koettgen et al.\cite{ceria02},
   and the approach convincingly explains the ionic conductivity of doped ceria obtained by experiment.
   Our previous research\cite{NAKATA2020109844} also follows the approach of Koettgen et al.\cite{ceria02}, with the method being used 
   to investigate the effect of Fe dopant on oxide ion diffusion in SrTiO$_{3-\delta}$.
   Here, we extend the FkMC approach to one based on machine learning (ML), i.e., FkMC-ML, in which the complex potential-energy 
   landscape is modeled in terms of ML-derived potentials, and the kMC simulation is performed 
   using these predicted changes in potential energy. 

   First, the  original FkMC is reviewed briefly and 
   its extension to ML-based potential correction is described in detail. 
   Second, the kinetics of hydride ion migration are evaluated using the new FkMC-ML model
   and the differences between  the original and the ML-corrected potentials  are discussed.
   We also compare the evaluated activation barrier of hydride ion with those obtained by experiment\cite{liu2019highly}
   to demonstrate the validity of the proposed approach.
   Third, the diffusion coefficient for the oxide ions is evaluated 
   and the effect of H concentration on both hydride and oxide ions is discussed, with the aim of understanding 
   the mechanism of ionic migration in oxyhydrides.
   Finally, the parallel-processing efficiency of FkMC-ML is evaluated to show the effectiveness of the approach.
    
\section{Theory and Method}

\subsection{Summary of the FkMC method}

    The kMC method is described in detail elsewhere\cite{kMCbook,kMCbasis01,kMCbasis02}. Here, 
    we briefly describe the kMC approach to simulating  ionic diffusion.  
    In the jump-diffusion kMC approach,
    the transition rate ($k_i^j$) from atom $i$ to atom $j$ 
    can be estimated as
    \begin{align}
      k_i^j = A 
          \mathrm{exp}
          \left[
            - \frac{E^\dagger_{ij}}{RT}
          \right],
      \label{RateCnst1}
    \end{align}
    where $E^\dagger_{ij}$, $R$, $T$, and $A$ are the activation energy, universal gas constant,
    temperature, and pre-exponential factor, respectively.
    Summing the respective transition rates gives an estimate of the total rate constant $r^{\mathrm{tot}}$, expressed as
    \begin{align}
      r^{\mathrm{tot}} 
      \label{transitionRate}
    = &
      \sum_{i}^{n^{\mathrm{atom}}}
      \sum_{j}^{n_i^{\mathrm{site}}}
      k_{i}^{j}, 
    \end{align}
    where $n^{\mathrm{atom}}$ and $n_i^{\mathrm{site}}$ are the total number of target atoms in the system and 
    number of nearest neighbor sites for atom $i$, respectively. For example, in the case of an oxide ion in perovskite crystal,
    $n_i^{\mathrm{site}} = 8$ for all oxygen sites.
    The transition probability from $i$ to $j$ ($p_i^j$)  is given by
    \begin{align}
      p_i^j =
      \label{kMCprobability}
      &   k_i^j / r^{\mathrm{tot}}.
    \end{align} 
    By iteratively selecting the transition from $i$ to $j$ based on the probability $p_i^j$, 
    the position of target site $i$ is updated.
    Likewise, the next $n+1$th step simulation time $t_{n+1}$ can be obtained 
    from current  time $t_{n}$ and the total rate constant in Eq.~\ref{transitionRate}: 
  \begin{align}
     t_{n+1}  = t_{n} - \frac{ln(r^\prime)}{r^{\mathrm{tot}}},
     \label{TimeStep}
  \end{align}
   where  $r^\prime$ is uniform random number from zero to one. 

    Because the total number of events is $n^{\mathrm{atom}} \times n^{\mathrm{site}}$,
    the total rate constant $r^{\mathrm{tot}}$ increases cubically
    with the size of lattice. Therefore, the simulation time step $t_{n+1}  - t_{n}$ 
    cubically decreases, and the kMC approach will be limited to relatively small systems.
    To reduce the computational cost of kMC,  
    several block separation schemes have been developed\cite{li2019openkmc,tada2013}, 
    aiming to achieve good parallel-processing efficiency.
    We have recently proposed an alternative type of parallelizable  scheme 
    involving atom-based fragmentation\cite{NAKATA2020109844},
    where the transition rate is estimated for each atom $i$
    and the maximum rate constant for an atom is defined as
  \begin{align}
    r_i             =&  \sum_{j \in n_i^{\mathrm{site}}} k_i^j,
  \\
    r^\mathrm{max}  =&  \mathrm{max} 
                        \left(
                          r_i
                        \right).
  \end{align}
  If we adopt dynamic renormalization\cite{kMCRenormalize},
  then the transition probability $p_i^j$ in  Eq.~\ref{kMCprobability}
  can be reformulated as
  \begin{align}
      p_i^j =
     \label{Prob02}
          \frac{r^\mathrm{max}}{r^{\mathrm{tot}}}
          \frac{k_i^j         }{r^\mathrm{max}}.
  \end{align}
  The transition event selection can then be partitioned into
  the selection of an atom with $r^\mathrm{max} / r^{\mathrm{tot}}$ and the selection of an event 
  for the independent atom $k_i^j / r^\mathrm{max}$.
  With this renormalization, several ions can be updated simultaneously ,
  and the parallelization of kMC becomes easier\cite{NAKATA2020109844}.
  In this study, we extend this approach by adopting ML potential correction.    
 
\subsection{Modification of transition rate with ML potential correction}
    
\begin{figure}
          \begin{center}
           \includegraphics[clip,width=8.0cm]{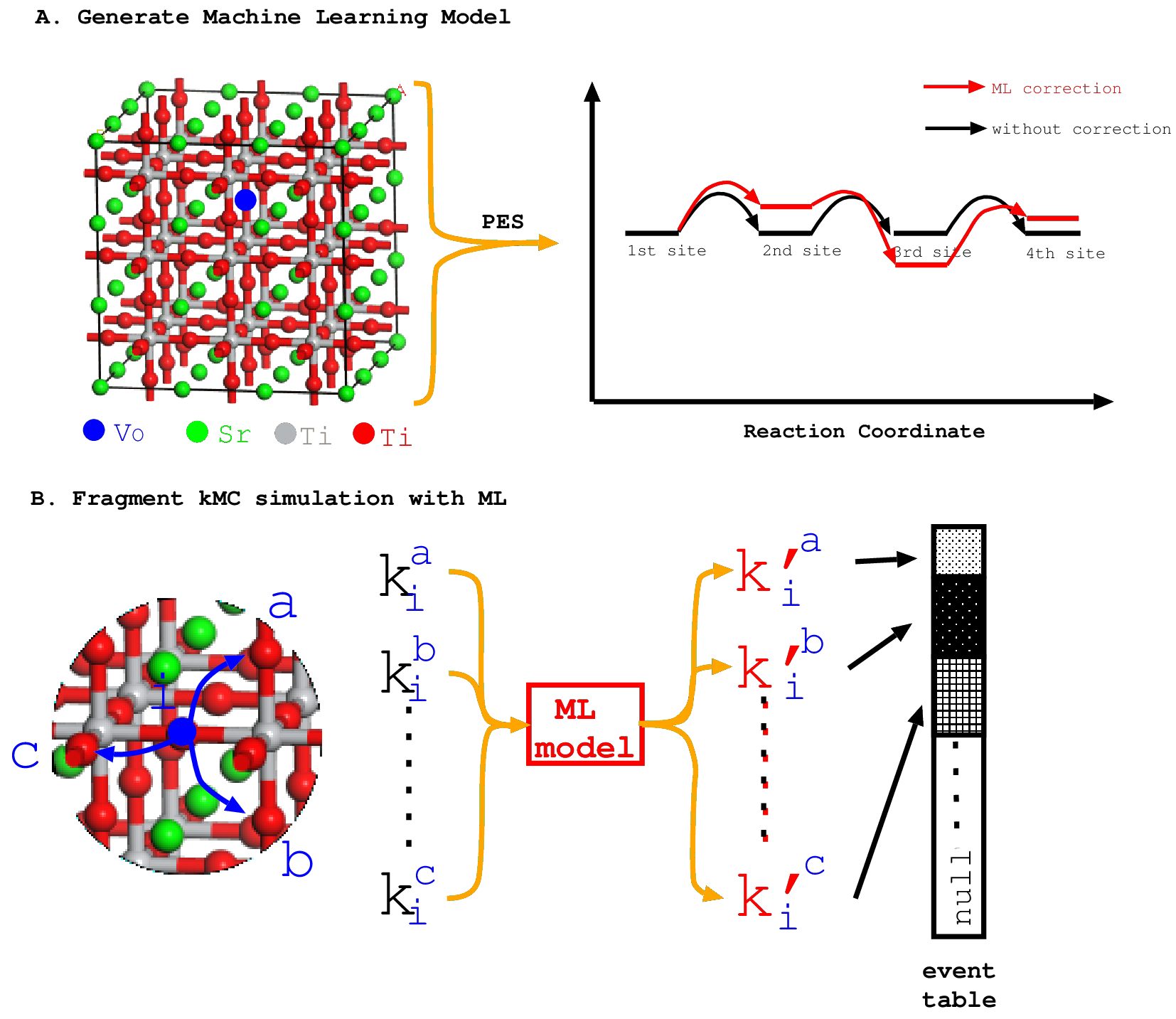} \\
          \end{center}
	      \caption{
              A: schematic illustration of the effect of PES correction using the ML model.
              B: schematic illustration integrating the ML model into kMC simulation.
              \label{Figure01}
		      }
	\end{figure}

    The key parameter for kMC simulation is the activation barrier $E^\dagger_{ij}$ from $i$  to $j$
    defined in Eq.~\ref{RateCnst1}. In the case of oxide ion migration in a single SrTiO$_3$ crystal,
    the PES is almost flat, as shown by black line in Fig.~\ref{Figure01}(a), 
    and no correction of the activation barrier of $E^\dagger_{ij}$ is required.

    However, the flat PES case is quite rare, with the actual PES often being affected by the presence of other types of ions. 
    The main idea in this study is to consider correcting PES changes by using an ML model (the red line in Fig.~\ref{Figure01}(a)).
    The transition rate $k_i^j$ can then be reformulated as 
     \begin{align}
       k_i^{j,\prime} = A 
           \mathrm{exp}
           \left[
             - \frac{E^\dagger_{ij} + \Delta E_{ij}^{\mathrm{ML}}/2}{RT} 
           \right],
       \label{newrate}
     \end{align}
     where $\Delta E_{ij}^{\mathrm{ML}}$ is the potential energy difference 
     between the final and the initial vacancy sites, which can be estimated via the ML model. 
     A similar reformulation of the transition rate can be found in the recent review of Koettgen et al.\cite{ceria02}. 

\begin{figure}
          \begin{center}
          \includegraphics[clip,width=8.0cm]{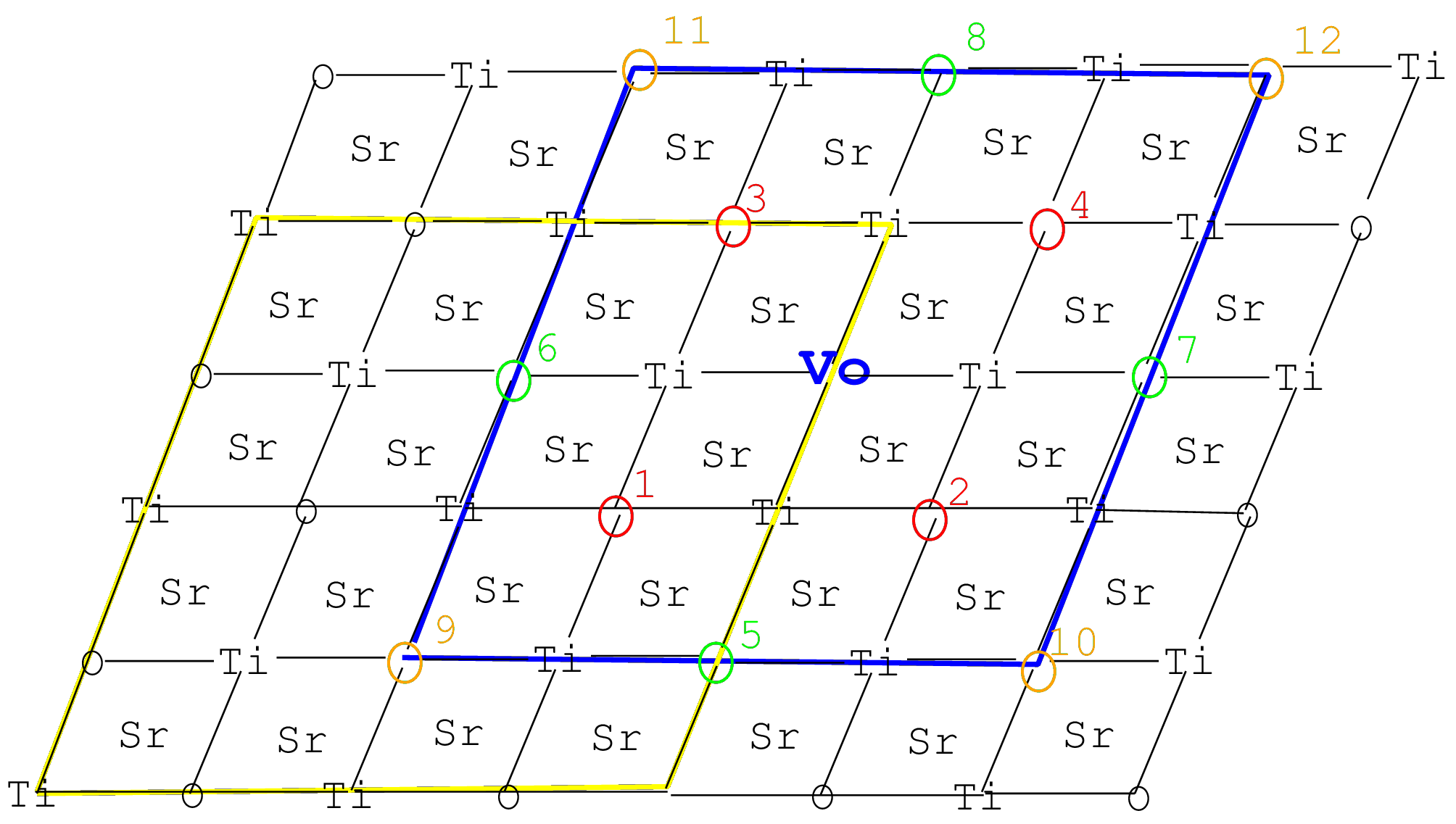} \\
          \end{center}
	      \caption{
              Schematic illustration of constructing the feature vector for oxygen vacancy $V_O$,
              with colored labels 1, 2, 3, $\cdots$ indicating the elements of the feature vector.
              \label{Figure02}
		      }
	\end{figure}

     As an example, for the case of a vacancy transition in an SrTiO$_3$ perovskite-type crystal, 
     the vacancy can jump to any of the eight nearest-neighbor sites (see Fig.~\ref{Figure01}(b),and Fig.~\ref{Figure02}).
     First, the transition rates from $i$ to the other sites ($a$, $b$, $\cdots$  $c$) are 
     set from a predefined simulation parameter. The transition rate $k_i^a$ is then refined 
     using the ML model
     \begin{align}
         \Delta E_{ij}^{\mathrm{ML}} =  f^\mathrm{ML} (\mathbf{X}),
         \label{MLmodel}
     \end{align}
     where $f^\mathrm{ML}$ is the ML model and $\mathbf{X}$ is the feature vector used to predict 
     $\Delta E_{ij}^{\mathrm{ML}}$. 
     In this study, the training of the ML model is aimed at developing the structure--energy relationship,
     with the energy being estimated using first-principles simulation.

     The feature vector is a one-dimension array that represents the atomic configuration in the crystal.
     To show how the vector $X$ is prepared, an example is shown in  Fig.~\ref{Figure02}).
     In this example, we are trying to construct the feature vector for oxygen vacancy $V_O$,   
     depicted in blue. The original position of $V_O$ is at the right edge of the periodic boundary condition (PBC) 
     (the yellow line in Fig.~\ref{Figure02}). We apply the minimum-image convention for $V_O$
     and the PBC lattice is shifted to the blue line to locate the $V_O$ at the center of the PBC (see Fig.~\ref{Figure02}).
     The atoms in the oxygen site are then reordered as the first nearest neighbor (red), 
     the second nearest neighbor (green), and the third nearest neighbor (orange). They are 
     labeled down to up and left to right (1, 2, 3, $\cdots$ 10 in  Fig.~\ref{Figure02}).  
     Using this labeling by the minimum image convention, the feature vector $X$ can be filled for any atomic species.
     In the case of SrTiO$_3$ perovskite-type crystal, all the oxygen sites are symmetrical,
     making it possible to use a single ML model $f^\mathrm{ML}$ to predict 
     the potential energy difference $\Delta E_{ij}^{\mathrm{ML}}$.
     By inserting Eq.~\ref{MLmodel} into Eq.~\ref{newrate},  $k_i^{j,\prime}$, $r_i$, and $r^\mathrm{max}$ can be updated
     and the transition probability $p_i^j$ can be updated using Eq.~\ref{Prob02}. 
     
\section{Computational details}

  As noted in the Introduction, we investigated the kinetics of hydride and oxygen ions with the aim of
  understanding how the hydride and oxygen ions interact with each other. 
  To achieve this, the FkMC-ML method was evaluated for four different levels of hydride ion concentration, namely for SrTiO$_{2.75}$H$_{0.25}$, 
  SrTiO$_{2.65}$H$_{0.35}$, and  SrTiO$_{2.55}$H$_{0.45}$, and  SrTiO$_{2.55}$H$_{0.60}$.
  The same concentrations of hydride ions (from 0.25 to 0.45) were used in the experimental 
  study of Liu et al\cite{liu2019highly}, with the simulated activation barriers being 
  compared with the experimental results. Although SrTiO$_{2.55}$H$_{0.60}$ was not used 
  in those experiments, we included it to evaluate fully how the simulation results 
  change with an increase in hydride ions. 
  The FkMC-ML simulation involves two main steps.
  The first step is to construct the ML model using the structure--energy relation
  and the second step is performing the  FkMC-ML simulation to predict diffusion coefficients.
  
  To construct the ML model, a first-principles simulation (DFT) was performed using 
  Quantum Espresso software\cite{qe01,qe02} with a
  Perdew-Burke-Ernzerhof functional\cite{PBE01,PBE02}.  
  We used ultrasoft pseudopotentials\cite{ultrasoft}
  and  the cutoff energy for the plane-wave basis set was taken to be 300 Ry. 
  The default convergence criterion of  1.0D-4 a.u  
  was used for the geometry optimizations, 
  and the default convergence  criteria 1.0D-6 a.u was used 
  for the self-consistent field calculations of the electronic states. 
  The system size of SrTiO$_3$ in the first-principles simulation was 3$\times$3$\times$3
  and the total number of atoms was 135.  
  (More detailed information about constructing an ML model is given in  
  the Results and Discussion section.)
  In this study, we used tensorflow\cite{tensorflow} to construct the neural network model, 
  where the neural network contained four hidden layers of 36 neurons each. 
  To prevent overfitting, Ridge regression (L2 normalization) was adopted, 
  with a coefficient of 0.001.

  In the second step, 
  a kMC-ML simulation was performed to evaluate the diffusion coefficients of the hydride ion and oxygen ion
  for each of the four hydride ion concentrations (SrTiO$_{2.75}$H$_{0.25}$, 
  SrTiO$_{2.65}$H$_{0.35}$, and  SrTiO$_{2.55}$H$_{0.45}$, and  SrTiO$_{2.55}$H$_{0.60}$).
  In this case, the simulation system size was 90$\times$90$\times$96 
  (the size of simulation system is 100 nm)   
  and 288,000,000 simulation steps were performed.
  To evaluate the apparent activation energies for the various hydride ion concentrations,
  the temperature range was set to 550 K, 600 K, 650 K, and 700 K, for which
  the diffusion coefficient  was experimentally measured\cite{liu2019highly}.
  The vacancy concentration was set to 0.1 \%.  The simulation was parallelized using 144 central processing units (CPUs).

  The FkMC-ML approach was implemented within the kMC program 
  (written in C++) and the program was parallelized by using a 
  message-passing interface.
  (The FkMC-ML program is available free of charge
  from GitHub (https://github.com/hiroyanakata/kMC.v02)).
  The parallel-processing efficiency of the kMC-ML was evaluated for the SrTiO$_{2.75}$H$_{0.25}$ case,
  where the system size was 900$\times$900$\times$900 (at a resolution of about 350 nm),
  and the kMC-ML simulation was performed using 1,000,000 steps.
  The computational costs were evaluated for the cases of 108, 216, 432, 864, and 1,000 CPUs 
  and the parallel-processing efficiency was evaluated.

\section{Results and discussion}

\subsection{Generating the ML model from the DFT dataset}

    The reference datasets for the ML model were prepared using a standard kMC simulation (i.e., without the ML potential correction)
    and 1,000 structures were randomly generated in the single kMC simulation run.
    The system size was 3$\times$3$\times$3 unit cells, as noted in the Computational Details section,
    and  independent kMC simulations were performed for four SrTiO$_{3-x}$H$_x$ cases, where $x$ was  0.25, 0.35, 0.45, or  0.60.
    For each simulation, one or two oxygen atoms were replaced with vacancies, giving eight sets of kMC 
    simulations.
    Note that the insertion of two oxygen vacancies is at a higher concentration than was used in the actual kMC 
    simulations, but we used both one-vacancy and two-vacancy models for the ML dataset to include the effect of the repulsion potential 
    between vacancies.  
    The size of the dataset  for evaluating the structure--energy relation was 8,000 and, using this dataset, 
    we constructed the ML model used to predict the PES in oxyhydrides.
 
    To construct the ML model, the feature vector $X$ in Eq.~\ref{MLmodel} is prepared using three steps.
    First, from the trajectory obtained by kMC, the selected transition event (V $\rightarrow$ O, V $\rightarrow$ H) 
    from site $i$ to site $j$ is determined and  the atomic labels are assigned using the minimum image convention 
    with respect to the center of the V$_O$ site ( site $i$)
    (see the Theory and Method section for details).
    Second, the labels of the atomic species (O, V, H) are replaced by the integer array (0, 1, 2)
    to obtain the sequential integer  vector $X = \left[1,0,0,2, \cdots, 2\right]$.
    To distinguish the transition event from the $i$ to $j$ site, we use the negative integer labels $-1$ and $-2$
    for the initial $i$ site and final $j$ site, respectively. 
    Finally, we introduce another integer labeling (using 1 and 2) to distinguish the hydride ion and oxygen ion
    transitions, with the transition label ``1'' or ``2'' being put at the top of the feature vector. 
    The completed feature vector then becomes $X = \left[1,-1,0,-2,2, \cdots, 2\right]$,
    which contains all the necessary information (reaction type, transition site positions, and 
    surrounding atomic species) to determine the change in PES.    
    
	\begin{figure}
          \begin{center}
          \includegraphics[width=8.0cm]{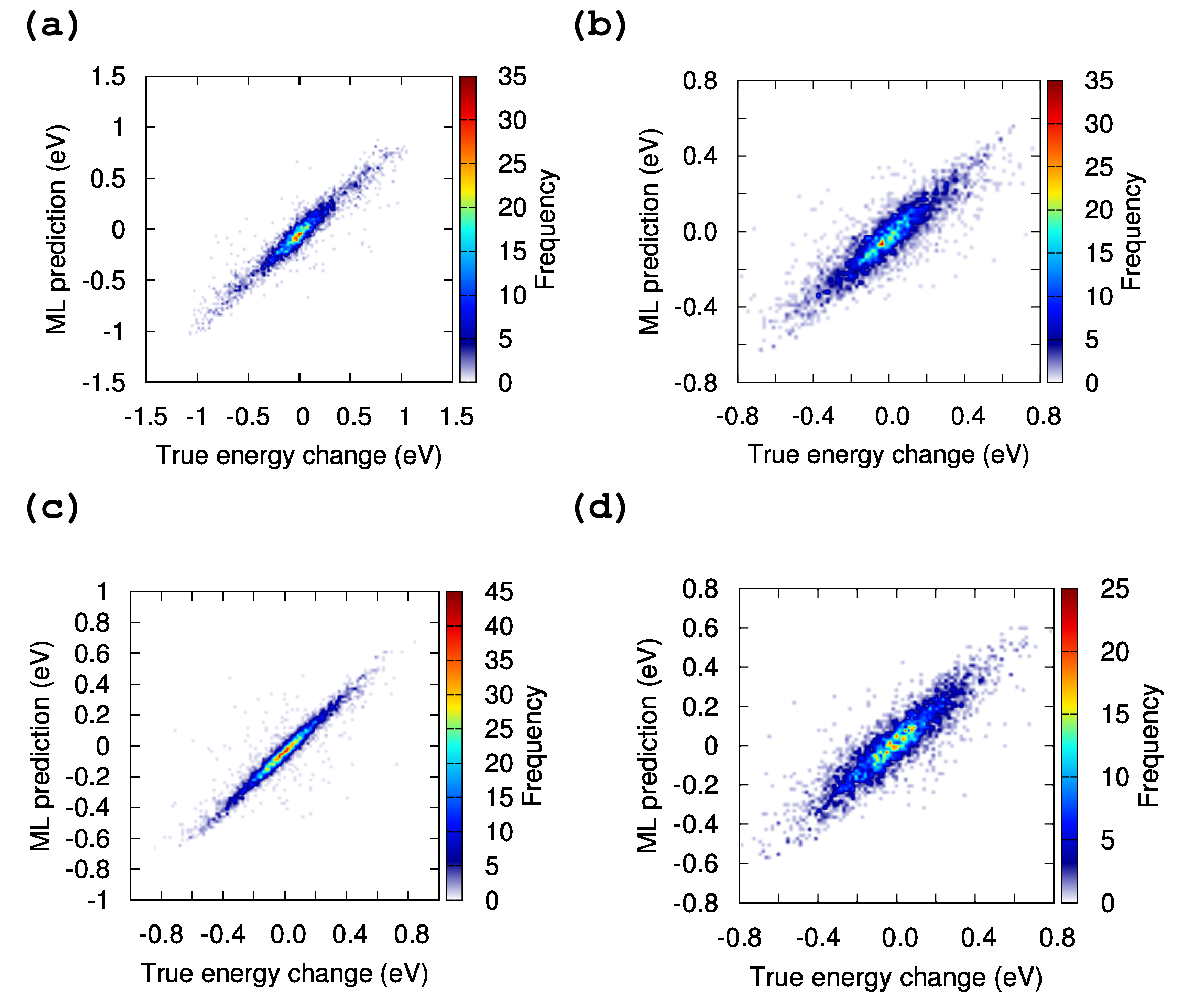} \\
          \end{center}
	      \caption{
              Comparison between DFT energy (horizontal axis) and ML prediction energy (vertical axis). 
              The colors show the numbers of data for each DFT energy:
              (a) the results for SrTiO$_{2.75}$H$_{0.25}$,
              (b) the results for SrTiO$_{2.65}$H$_{0.35}$,
              (c) the results for SrTiO$_{2.55}$H$_{0.45}$, and
              (d) the results for SrTiO$_{2.40}$H$_{0.60}$.
              \label{Figure03}
		      }
	\end{figure}

    The comparison between the ML energy (energy predicted by the ML model) and 
    the density functional theory (DFT) energy (energy estimated by the first-principles simulation)
    is depicted in Fig.~\ref{Figure03}.
    In Fig.~\ref{Figure03}, the horizontal axis represents the DFT energy 
    and the vertical axis represents the predicted ML energy. The DFT energy denotes
    the energy difference in the ionic transition from the $i$ site to the $j$ site.
    As shown in Fig.~\ref{Figure03}, the ML model reproduces the DFT energy   
    quite well and the root mean square error in Eq.~\ref{newrate} is 0.04 eV, 
    which is less than the error expected for experimentally determined activation barriers\cite{liu2019highly}. 
    From Fig.~\ref{Figure03}, we note that the most of the DFT energies are located 
    in a vicinity of zero, indicating that the PES is nearly flat. 
    Furthermore,  the comparison between ML and DFT energy shows a similar trend toward positive 
    and negative changes in PES. These results suggest that the ML model developed
    is adequate for the kinetic simulation of oxyhydrides.
    We therefore adopt this ML model for predicting the diffusion coefficients of hydride and oxide ions in SrTiO$_{3-\delta}$.

\subsection{Hydride ion diffusion coefficient evaluation using FkMC-ML}

\begin{figure}
          \begin{center}
          \includegraphics[clip,width=8.0cm]{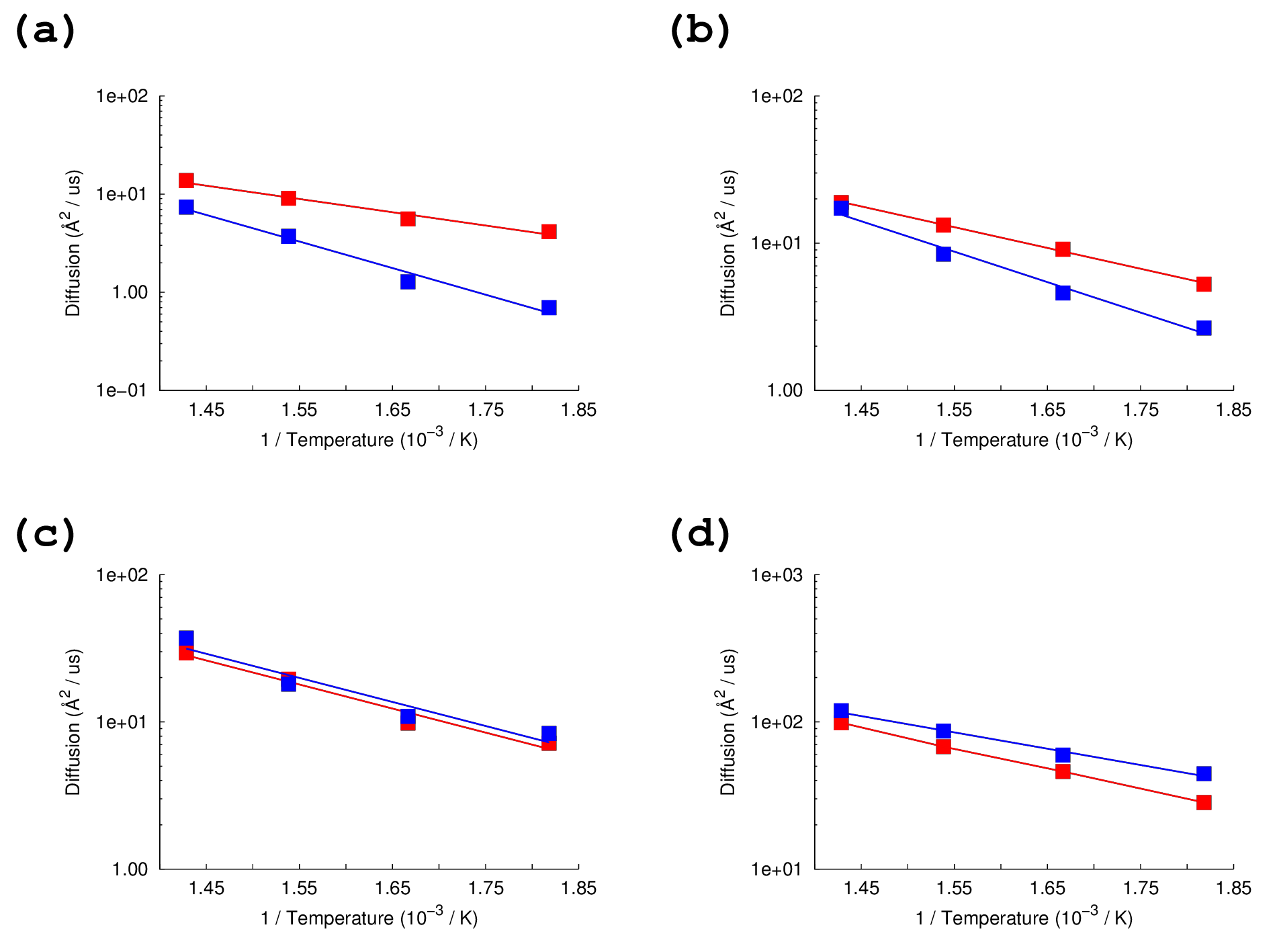} \\
          \end{center}
	      \caption{
              The diffusion coefficient for hydride ions, with the red closed squares denoting the diffusion coefficients using kMC-ML,
              and the blue closed squares denoting those using kMC. The horizontal axis represents the inverse of temperature and the vertical axis 
              represents the diffusion coefficients:
              (a) the results for SrTiO$_{2.75}$H$_{0.25}$,
              (b) the results for SrTiO$_{2.65}$H$_{0.35}$,
              (c) the results for SrTiO$_{2.55}$H$_{0.45}$, and
              (d) the results for SrTiO$_{2.40}$H$_{0.60}$.
              \label{Figure04}
		      }
	\end{figure}

   The hydride ion diffusion coefficients were evaluated for four
   oxyhydrides (SrTiO$_{3-x}$H$_{x}$, where x= 0.25, 0.35, 0.45, and 0.60), as noted in the Computational Details section. 
  To evaluate the effect of ML correction on the diffusion coefficients, results for the standard kMC (i.e., without ML) 
  were obtained and these results were compared with those obtained with the help of ML potentials.
  The results for the diffusion coefficients are shown in Fig.~\ref{Figure04},
  where there is a significant difference between the diffusion coefficients obtained with and without the ML correction.
  For hydride ion concentrations of  $x=0.25$ or $0.35$, the PES accelerates the diffusion of hydride ions 
  significantly, whereas the diffusion coefficient using kMC-ML remains the same for $x=0.45$ 
  and is slightly reduced for $x=0.60$.

  We can also note the difference in activation energy with and without PES correction.
  A summary of the activation barriers with and without ML correction
  is shown in Table~\ref{TABLE01}. 
  Without the effect of PES correction, the activation barrier monotonically decreased with an increase 
  in the concentration of hydride ions. The largest activation barrier was 0.53 eV for  x=0.25, 
  and the smallest was 0.27 eV. 
  The changes of activation barriers  can be understood in terms of slow oxygen diffusion. 
  When the hydride ion concentration is small, the diffusion of hydride ions is limited 
  by the surrounding oxygen ions. Therefore, the diffusion of hydride ions is affected by the oxygen ion concentration, 
  with the activation barrier increasing with oxygen ion concentration.    
  However, without PES correction, the estimated activation barriers are slightly larger 
  than those obtained experimentally (See  Table~\ref{TABLE01}).  
  Inclusion of ML correction in Eq.~\ref{newrate} changes the activation barrier from 0.42 eV to 0.28 eV for $x=0.35$ 
  and this simulation result with ML correction agrees well with the experimentally derived activation barrier (0.28 eV).	
  Likewise, for $x=0.45$, the simulated activation barrier with ML is 0.32 eV, agreeing well with experiment (0.30 eV), 
  which suggests that the simulation results are reasonable. 

\begin{table}[h!]
\caption[]{
     Apparent activation barrier for hydride ion diffusion in SrTiO$_{3-x}$H$_x$,
    for concentrations x= 0.25, 0.35, 0.45, and 0.60. 
    ML and non-ML denote the results obtained by kMC-ML and standard kMC, respectively.
    N/A indicates that results are not available.}
\label{TABLE01}
\begin{tabular}{lrrr}\hline
         x            &   non-ML           &    ML   &   Liu et al.\cite{liu2019highly}     \\\hline
     0.25             &  0.54             &  0.27   &   N/A                           \\
     0.35             &  0.42             &  0.28   &   0.28                           \\
     0.45             &  0.32             &  0.32   &   0.30                           \\     
     0.60             &  0.22             &  0.27   &   N/A                           \\\hline
\end{tabular}
\\
\end{table}%

  Note that the activation barriers estimated via FkMC-ML do not change very much with the variation of hydride ion concentration.
  However, without ML correction, the activation barrier is significantly changed, as noted above, indicating the importance of the ML potential contribution. 
  For the case of $x=0.25$, the activation barrier of 0.54 eV is close to the oxygen migration barrier
  in SrTiO$_3$ (0.6 eV), which suggests that the rate-determining step for hydride ions is oxygen migration.    
  Using ML potentials decreases the barrier to 0.27 eV, which suggest that the inclusion of the PES  
  may  affect not only hydride ion migration but also oxygen ion migration, 
  with both hydride and oxygen ion diffusion being accelerated. Because the hydride and oxygen ions interact significantly,
  the diffusion of oxygen should be analyzed to understand fully how hydride ion diffuse in SrTiO$_3$ crystals. We therefore also
  analyzed the kinetics of oxygen diffusion.
 
\subsection{Oxygen ion diffusion coefficients}

\begin{figure}
          \begin{center}
          \includegraphics[clip,width=8.0cm]{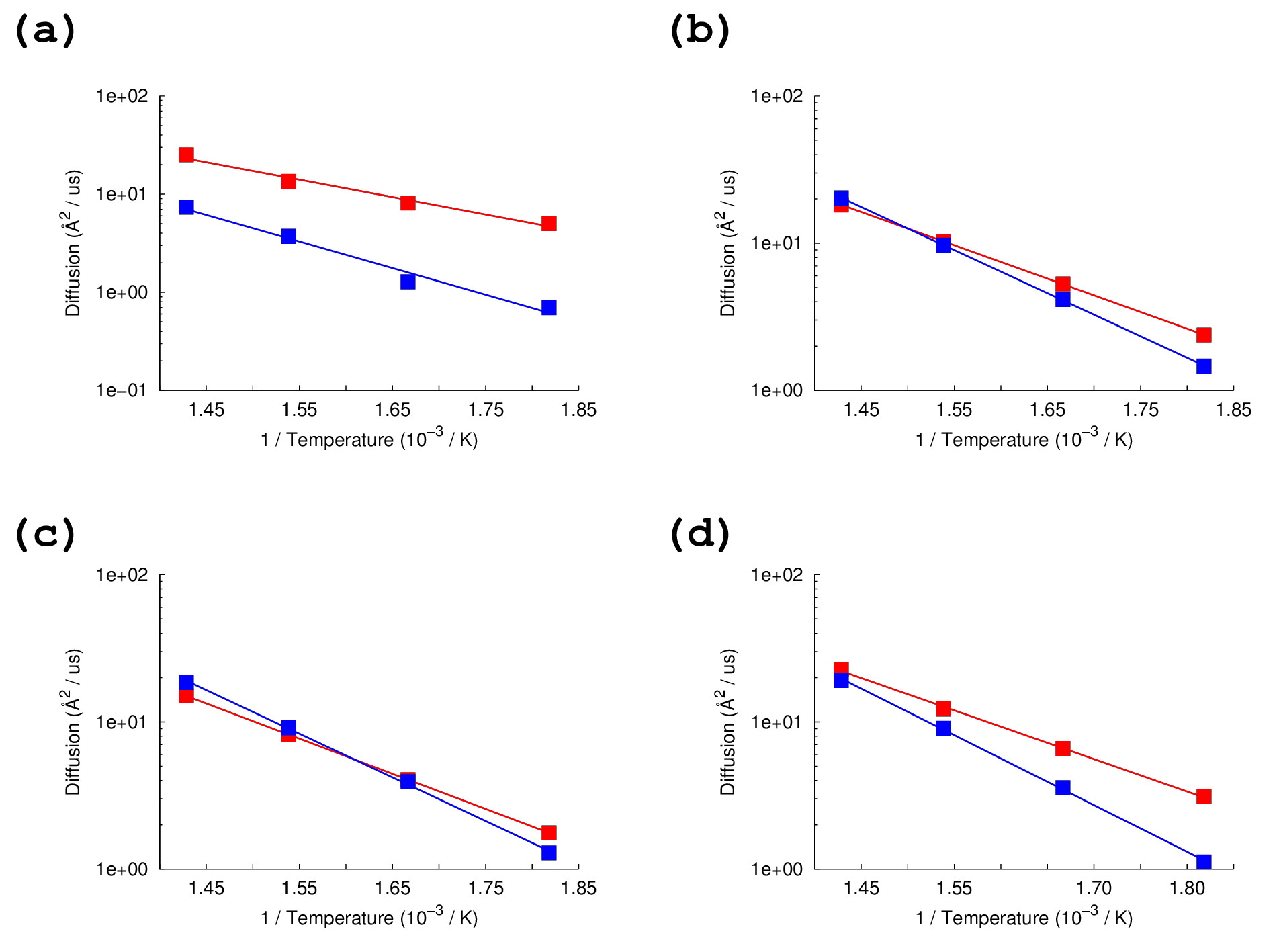} \\
          \end{center}
	      \caption{
              The diffusion coefficient for oxygen ions, with the red closed square denoting the diffusion coefficients using kMC-ML,
              and the blue closed square denoting those using kMC. The horizontal axis is inverse of temperature, and the vertical axis 
              represents the diffusion coefficients:
              (a) the results for SrTiO$_{2.75}$H$_{0.25}$,
              (b) the results for SrTiO$_{2.65}$H$_{0.35}$,
              (c) the results for SrTiO$_{2.55}$H$_{0.45}$, and
              (d) the results for SrTiO$_{2.40}$H$_{0.60}$.
              \label{Figure05}
		      }
	\end{figure}

  To understand fully the kinetics of SrTiO$_3$ oxyhydride,  oxygen ion diffusion was also evaluated for the 
  four hydride ion concentrations.  The results for the diffusion coefficients are shown 
  in Fig.~\ref{Figure05} and the activation energies are listed in Table~\ref{TABLE02}. 
  Without PES correction, the activation barriers for oxygen ions do not depend on the 
   concentration of hydride ions. The activation barriers are around 0.6 eV, which is 
  the same as for oxygen migration in a single SrTiO$_3$ crystal.

  Inclusion of the ML potential contribution changes significantly the kinetics of oxygen diffusion,
  particularly for small hydride ion concentrations.
  When the hydride ion concentration is 0.25, the diffusion coefficient for oxygen
  using kMC-ML is ten times larger than that obtained without using ML potentials. 
  The contribution of ML potentials to the diffusion coefficient is less noticeable 
  when the simulation temperature is high, which is reasonable 
  because the small changes in potential can be neglected given the increases from thermal fluctuation. 
  Because of the temperature dependence, the apparent activation energy of oxygen ion diffusion 
  decreases from 0.54 eV to 0.35 eV for $x=0.25$. Similar trends can be observed in the simulation results for 
  the  other hydride ion concentrations, with the apparent activation energies being decreased 
  by around 0.15 eV using the ML potential contributions for each hydride ion concentration. 
   
\begin{table}[h!]
\caption[]{
     Apparent activation barrier for oxygen ion diffusion in SrTiO$_{3-x}$H$_x$,
    for concentrations x= 0.25, 0.35, 0.45, and 0.60. 
    ML and non-ML denote the results obtained by kMC-ML and standard kMC, respectively.}
\label{TABLE02}
\begin{tabular}{lrrr}\hline
         x            &   non-ML           &    ML    \\\hline
     0.25             &  0.54             &  0.35    \\
     0.35             &  0.58             &  0.45    \\
     0.45             &  0.59             &  0.47    \\     
     0.60             &  0.63             &  0.43    \\\hline
\end{tabular}            \\                  
\end{table}%

\subsection{Hydride and oxygen ion diffusion mechanisms}

\begin{figure}
          \begin{center}
          \includegraphics[clip,width=8.0cm]{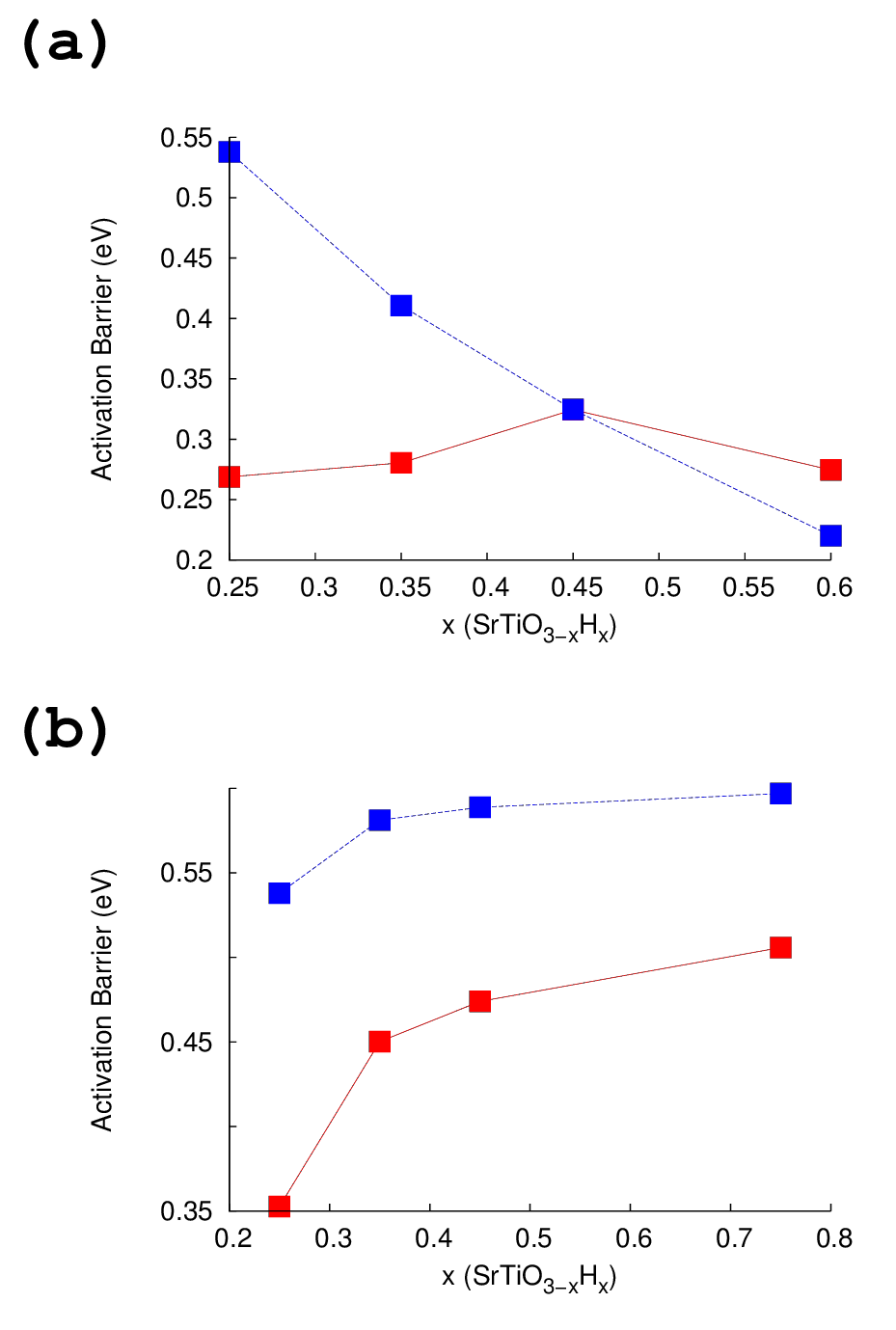} \\
          \end{center}
	      \caption{
              Activation barriers for hydride ions in SrTiO$_{3-x}$H$_{x}$, for concentrations $x=$ 0.25, 0.35, 045, and 0.6.
              The red closed squares denote the diffusion coefficients using kMC-ML
              and the blue closed squares denote those using kMC:
              (a) the activation barrier for hydride ion diffusion and
              (b) the activation barrier for oxygen  ion diffusion. 
              \label{Figure06}
		      }
	\end{figure}

  To help understand the diffusion of hydride and oxygen ions,
  the changes in the activation energy for changing concentrations of hydride ions 
  are shown in Fig.~\ref{Figure06}. 
  In this figure,
  the difference between the blue and red lines shows the impact of the PES correction,
  summarizing the impact of interactions between hydride and oxygen ions.

  The activation barrier for oxygen ions is uniformly reduced by considering PES changes, 
  with the activation energy being around 0.4 eV (see Fig.~\ref{Figure06}(b)). 
  By contrast, with hydride ion diffusion,  the effect of PES correction does depend on 
  the concentration of hydride ions. There is a reduction in activation energy for small hydride 
  ion concentrations ($x=0.25$ or $0.35$), whereas the activation energy of the hydride ions remains similar for a higher
  hydride ion concentration ($x=0.45$).

  At a low hydride ion concentration ($x=0.25$), hydride ions cannot diffuse 
  without involving the diffusion of oxygen ions.  Because the estimated oxygen diffusion is accelerated when considering ML potentials, 
  the estimated diffusion of hydride ions will also be accelerated significantly. 
  For a moderately high hydride ion concentration ($x=0.35$ or  $0.45$), 
  hydride ions can diffuse while interacting mainly with other hydride ions. The activation barrier without considering ML potentials will then
  also become less dependent on the activation barriers for oxygen ions (0.42 eV or 0.32 eV), confirming that slow oxygen diffusion 
  is not the primary rate-determining factor in such cases. Therefore, considering the decreased oxygen activation barrier 
  by using ML potentials does not significantly affect the activation barriers for hydride ions.
  Likewise, at a high hydride ion concentration ($x=0.60$), a decreased oxygen activation barrier 
  would not accelerate hydride ion diffusion. In contrast to the other cases, using the ML potentials
  suppresses the diffusion of hydride ions, and the activation barrier is observed to increase from 0.22 eV to 0.27 eV.   
  In summary, the activation barrier for hydride ions is around 0.3 eV, independent of the hydride ion concentration.
    
  In this section, we have investigated the hydride and oxygen ion diffusion mechanisms
  by analyzing the activation barriers' dependence on the hydride ion concentration.
  Considering the inhomogeneous PES created via with ML model, the simulation results concur with
  the experimental activation barriers obtained by Liu et al.\cite{liu2019highly} (see Table~\ref{TABLE01}).  
  This offers an insight into why the experimentally determined activation barrier does not 
  depend on the hydride ion concentration.  The simulation results indicate that there are 
  two factors determining hydride ion diffusion, namely that increasing the concentration of hydride ions 
  accelerates hydride ion diffusion and that using an inhomogeneous PES accelerates the diffusion of oxygen. 
  These factors affect the rate-determining step, particularly at low hydride ion concentrations.  

\subsection{Parallel-processing efficiency}

\begin{figure}
          \begin{center}
          \includegraphics[clip,width=8.0cm]{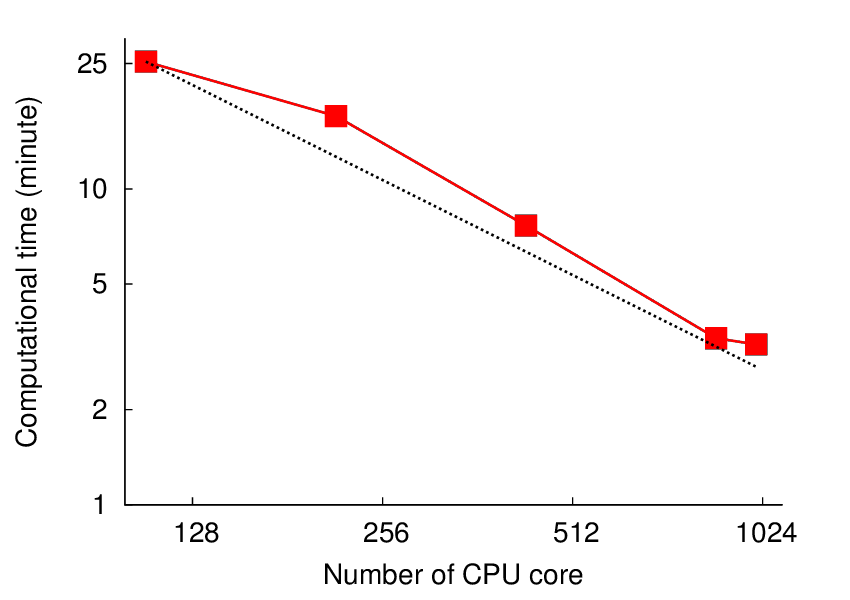} \\
          \end{center}
	      \caption{
              Computational timing and its parallel-processing efficiency for FkMC-ML.
              The black dashed line is the ideal computational time,
              estimated by using the computational time with 64 cores. 
              \label{Figure07}
		      }
	\end{figure}

  Finally, the parallel-processing efficiency of FkMC-ML was evaluated 
  for various numbers of CPUs in the range of 64 to 1,000, as noted in the Computational Details section. 
  The results are shown in Fig.~\ref{Figure07}. 
  In the figure, the red line is the actual computation time, 
  and the black line is the ideally parallelized performance.
  Ideally, the computational time should decrease as the number of CPUs increases.
  We obtained a parallel efficiency of 84.92 \% using 1,000 CPUs in a comparison with the 108-CPU case.
  The effective parallel-processing efficiency (the ratio by which the program can be parallelized) 
  was 99.979 \%, suggesting that most aspects of the simulation had been parallelized.

\section{Conclusions}

  In this study, we have investigated hydride and oxygen ion diffusion 
  in SrTiO$_3$ oxyhydride crystals using a FkMC-based simulation.
  Experimentally obtained observations about the interaction between oxygen ion and hydride ion diffusion
  were also observed in our simulations, with our results 
  concurring with the experimentally obtained activation barriers for hydride ion diffusion.
  
  The FkMC simulation indicates that the activation barrier for hydride ions  
  is affected by slow oxygen diffusion and that the estimates for oxygen diffusion rates
  are increased when considering an inhomogeneous PES. 
  Therefore, the inclusion of such PES changes should be an important aspect of meaningful FkMC models.
 
  In this study, we have developed an FkMC-based approach that includes neural-network capabilities,
  which enables the complex PES landscape to be included easily in the simulation model.
  The parallel-processing efficiency of the proposed approach is promising, suggesting that 
  the approach can be widely used for simulating ionic diffusion in crystals.
  We hope that our FkMC-ML simulation will aid the understanding 
  of ion diffusion mechanisms in crystals.
   
\section*{ACKNOWLEDGMENTS}
We thank 
the Research Institute for Information Technology
 at Kyushu University for providing computational resources. 
This research also used the computational resources of 
the Fujitsu PRIMERGY CX400M1/CX2550M5(Oakbridge-CX) at the
Information Technology Center of the University of Tokyo
through the HPCI System Research project (Project ID:hp200015).

\bibliographystyle{aip}
\bibliography{FkMC}

\newpage

\end{document}